\documentclass[a4paper]{revtex4}
\usepackage{epsfig}
\usepackage{amsfonts}
\usepackage{amsmath}
\usepackage{graphicx}

\newcommand{\eq}{\begin{equation}}
\newcommand{\eqx}{\end{equation}}
\newcommand{\eqn}{\begin{eqnarray}}
\newcommand{\eqnx}{\end{eqnarray}}
\newcommand{\f}[2]{\frac{#1}{#2}}

\newcommand{\al}{\alpha}
\newcommand{\bt}{\beta}

\newcommand{\eps}{\varepsilon}
\newcommand{\nn}{{\cal N}}

\newcommand{\tr}{\mbox{\rm tr}\,}
\newcommand{\li}{\mbox{\rm li}_2}

\newcommand{\OO}[1]{{\cal O}\left(#1\right)}
\newcommand{\taubol}{\tau_\Pi^{Boltzmann}}
\newcommand{\btl}{\tilde{b}}
\newcommand{\rsq}{{\mathfrak R}^2}
\newcommand{\qqqq}{\quad\quad\quad\quad}

\newcommand{\cor}[1]{\left\langle{#1}\right\rangle}

\begin{document}

\title{Viscous hydrodynamics relaxation time from AdS/CFT}  

\author{Micha\l\ P. Heller}\email{heller@th.if.uj.edu.pl}
\affiliation{Institute of Physics, Jagellonian University,
Reymonta 4, 30-059 Krakow, Poland.}
\author{Romuald A. Janik}\email{ufrjanik@if.uj.edu.pl}
\affiliation{Institute of Physics and M. Kac Complex Systems Research
  Centre, Jagellonian University,
Reymonta 4, 30-059 Krakow, Poland.}

\begin{abstract}
We consider an expanding boost-invariant plasma at strong coupling
using the AdS/CFT correspondence for $\nn=4$ SYM.
We determine the relaxation time in second order viscous hydrodynamics
and find that it is around thirty times shorter than 
weak coupling expectations. We find that the
nonsingularity of the dual geometry in the string frame necessitates
turning on the dilaton which leads to a nonvanishing expectation value
for $\tr F^2$ behaving like $\tau^{-10/3}$.  
\end{abstract}

\maketitle

\section{Introduction}

It is believed that the quark-gluon plasma (QGP) produced at RHIC in
heavy-ion collisions is strongly coupled (see e.g. \cite{review}) and
is well described by almost perfect fluid hydrodynamics \cite{hydro}. 
Therefore it is very
interesting to develop methods for studying its properties
nonperturbatively from first principles. A very powerful tool in the
study of the dynamics of gauge theory at strong coupling is the
AdS/CFT correspondence \cite{adscft}. Although so far there does not
exist a version of the correspondence with a gauge theory which would
have all the features of QCD, even the 
simplest version for $\nn=4$ Super Yang-Mills theory was argued to
share a lot of properties {\em at finite temperature} with QCD
plasma. But of course one has to keep in mind the differences which
may not be important for some features of the dynamics but which may
be crucial for other phenomenae. In this paper we thus work
exclusively with hot expanding plasma in the $\nn=4$ SYM theory.

Extensive work has been done in the understanding of transport
properties of the plasma at fixed temperature like calculating the
shear viscosity \cite{son,other}. Much less is known about the
properties of more dynamical time-dependent processes.
On a qualitative level, thermalization has been
suggested to correspond to black hole formation in
the bulk of the five-dimensional dual geometry
\cite{nastase}, while cooling was advocated to correspond to black
hole motion in the 5th direction \cite{zahed}.

In \cite{JP1} a quantitative framework has been proposed for studying
the time-dependent expansion of a boost-invariant plasma system. The
criterion of nonsingularity of the dual geometry was shown to predict
almost perfect fluid hydrodynamic expansion \cite{JP1} with leading
deviations coming from shear viscosity \cite{RJ} with the shear
viscosity coefficient being exactly equal to the one derived in the
static case in \cite{son}. Further work in this framework include
\cite{SJSIN,JP2,BAK,SJSIN2,Kajantie}. The aim of this paper is to
investigate in more detail the hydrodynamic expansion and to determine
the remaining parameter in second order viscous hydrodynamics
\cite{IS,BRW} -- the
relaxation time $\tau_\Pi$. This requires going one order higher in
the subasymptotic expansion of the geometry.   

The plan of this paper is as follows. In section II we will describe
the kinematic regime of longitudinal boost invariance. Then in section
III we will briefly review second order viscous hydrodynamics. In
section IV we describe the AdS/CFT methods used here and review, in the
following section, the results obtained so far and a method of
determining the relaxation time $\tau_\Pi$. In section VI we give
final results for $\tau_\Pi$ and in section VII we analyze the
incorporation of the dilaton and calculate the expectation value of
$\tr F^2$. We close the paper with a discussion.

\section{Boost invariant kinematics}

An interesting kinematical regime of the expanding plasma is the
so-called central rapidity region. There, as was suggested by Bjorken
\cite{Bjorken}, 
one assumes that the system is invariant under longitudinal
boosts. This assumption is in fact commonly used in realistic
hydrodynamic simulations of QGP \cite{hydro}. If in addition we assume
no dependence on transverse coordinates (a limit of infinitely large
nuclei) the dynamics simplifies enormously.

In order to study boost-invariant plasma configurations it is
convenient to pass from Minkowski coordinates $(x^0,x^1, x_\perp)$ to
proper-time/spacetime rapidity ones $(\tau,y,x_\perp)$ through
\eq
x^0=\tau \cosh y \qqqq x^1=\tau \sinh y
\eqx
The object of this work is to describe the spacetime dependence of the
energy-momentum tensor of a boost-invariant plasma in $\nn=4$ SYM
theory at strong coupling. The symmetries of the problem reduce the
number of independent components of $T_{\mu\nu}$ to three.
Energy-momentum conservation $\partial_\mu T^{\mu\nu}=0$ and
tracelessness $T^\mu_\mu=0$ allows to express all components in
terms of just a single function -- the energy density $\eps(\tau)$ in the
local rest frame.
Explicitly we have \cite{JP1}
\eqn
T_{\tau\tau} &=& \eps(\tau) \\
T_{yy} &=& -\tau^2 \left( \eps(\tau)+\f{d}{d\tau} \eps(\tau) \right) \\
T_{xx} &=& \eps(\tau)+\f{1}{2} \tau \f{d}{d\tau} \eps(\tau)
\eqnx
Gauge theory dynamics should now pick out a definite function
$\eps(\tau)$. The aim of this paper is to find its behaviour at large
proper-times, up to subsubasymptotic terms, and to interpret this
behaviour in terms of parameters of second order hydrodynamics which
we will describe in the next section.

\section{Second order viscous hydrodynamics}

The object of a hydrodynamic model is to determine the spacetime
dependence of the energy-momentum tensor for an expanding (plasma)
system.
The simplest dynamical assumption is that of a perfect fluid. This amounts to
assuming that the energy momentum has the form
\eq
\label{e.perfect}
T_{\mu\nu}=(\eps+p)u_\mu u_\nu+p \eta_{\mu\nu}
\eqx 
where $u^\mu$ is the local 4-velocity of the fluid ($u^2=-1$), $\eps$
is the energy density and $p$ is the pressure. In the case of $\nn=4$
SYM theory that we consider here $T^\mu_\mu=0$ and hence $\eps=3p$.
The equation of motion that one obtains from energy conservation in
the boost-invariant setup is
\eq
\partial_\tau \eps = -\f{\eps+p}{\tau} \equiv -\f{4}{3} \f{\eps}{\tau}
\eqx 
whose solution is the celebrated Bjorken result
\eq
\eps=\f{1}{\tau^{\f{4}{3}}}
\eqx
Once one wants to include dissipative effects coming from shear
viscosity, the description becomes more complex. In a first
approximation one adds to the perfect fluid tensor a dissipative
contribution $\eta (\nabla_\mu u_\nu+\nabla_\nu u_\mu)$, where $\eta$
is the shear viscosity of the fluid. The resulting equations of motion
get modified to 
\eq
\partial_\tau \eps = -\f{4}{3} \f{\eps}{\tau} +\f{4\eta}{3\tau^2}
\eqx 
Note that in the above equation the shear viscosity is generically
temperature dependent ($\eta \propto T^3$ in the $\nn=4$ case) and
hence $\tau$ dependent. In order to have a closed system of equations
we have to incorporate this dependence through
\eq
\label{e.etaeps}
\eta =A \cdot \eps^{\f{3}{4}}
\eqx
with $A$ being some numerical coefficient. 

However this so-called first order formalism suffers from a number of
problems. Firstly it is inconsistent with relativistic invariance
(causality) - excitations may propagate at speeds faster than
light. Secondly these equations suffer from some unphysical behaviour
(see e.g. \cite{BRW}).

In order to cure these problems Israel and Stewart introduced a second
order theory \cite{IS} with an additional parameter - the
so-called relaxation time $\tau_\Pi$ which can overcome the problems
with causality. This theory has found applications in modelling
heavy-ion collisions \cite{BRW1}. The corresponding equations of
motion, again in the Bjorken regime which we are considering here, are
now
\eqn
\label{e.i}
\partial_\tau \eps &=& -\f{4}{3} \f{\eps}{\tau} +\f{\Phi}{\tau} \\
\label{e.ii}
\tau_\Pi \partial_\tau \Phi &=& -\Phi +\f{4\eta}{3\tau^2}
\eqnx
where $\Phi$ is related to the dissipative part of the energy momentum
tensor $\Pi_{\mu\nu}$ through $\Phi=-\tau^2 \Pi^{yy}$.
Let us note that when $\tau_\Pi \to 0$ the equations reduce to the
first order formalism case. In the above equations one has to keep in
mind that the shear viscosity $\eta$ is again given by
(\ref{e.etaeps}). Hence the only remaining independent parameter is therefore
$\tau_\Pi$.

In the derivation of the second order viscous hydrodynamics from
Boltzmann equations one finds that the ratio of $\tau_\Pi$ to $\eta$
is given in terms of the pressure $p$ \cite{BRW}
\eq
\label{e.taubol}
\taubol=2\beta_2 \eta=\f{3\eta}{2p}
\eqx 
This is the value usually used in viscous hydrodynamic
simulations. Tha aim of this work is to determine the relaxation time
in the strong coupling regime using the AdS/CFT correspondence. To
this end let us parametrize the relaxation time at strong coupling as
\eq
\tau_\Pi= r \cdot \taubol
\eqx
where $\taubol$ is {\em defined} by the expression (\ref{e.taubol})
and $r$ is a numerical coefficient. The aim of this paper is to
determine $r$.

\section{AdS/CFT description of an expanding boost-invariant plasma}

The procedure adopted in \cite{JP1} to describe, using AdS/CFT, an
expanding system of plasma in $\nn=4$ SYM is (i) consider a
family of possible behaviours of the spacetime expectation value of
the energy-momentum tensor $\cor{T_{\mu\nu}}$, (ii) for each of those
$\cor{T_{\mu\nu}}$'s find the dual geometry which generically will be
singular and (iii) use the criterion of nonsingularity of the
constructed geometry to pick out the {\em physical} spacetime profile of
$\cor{T_{\mu\nu}}$.

For the plasma configuration considered in this paper, as described in
section II, all possible possible boost-invariant profiles of
$\cor{T_{\mu\nu}}$ can be expressed in terms of a single function
$\eps(\tau)$ which is just the energy density in the local rest frame.

The construction of a dual geometry then proceeds as follows
\cite{Skenderis}.  First we adopt the
Fefferman-Graham coordinates \cite{fg} for the 5-dimensional metric
\eq
ds^2=\f{\tilde{g}_{\mu\nu} dx^\mu dx^\nu + dz^2}{z^2}
\eqx
where the $z$ coordinate is the `fifth' coordinate while $\mu$ is a 4D
index. $\tilde{g}_{\mu\nu}$ is here a function both of the 4D
spacetime coordinates {\em and} of the `fifth' coordinate $z$.
Then we have to solve Einstein equations with negative cosmological
constant:
\eq
\label{e.einst}
E_{\al\bt} \equiv R_{\al\bt}-\f{1}{2} g_{\al\bt} R-6g_{\al\bt}=0
\eqx
with a boundary condition for $\tilde{g}_{\mu\nu}$ around $z=0$:
\eq
\label{e.bc}
\tilde{g}_{\mu\nu}=\eta_{\mu\nu}+z^4 \tilde{g}^{(4)}_{\mu\nu}+\ldots
\eqx
where the fourth order term is related to the expectation value of the
energy-momentum tensor through 
\eq
\label{e.adstmunu}
\cor{T_{\mu\nu}} = \f{N_c^2}{2\pi^2} g_{\mu\nu}^{(4)}
\eqx
It is convenient to ignore the factor $N_c^2/(2\pi^2)$ throughout the
calculation and only reinstate it in the final result when
e.g. writing the energy density or shear viscosity directly in terms
of the temperature.

For the boost-invariant geometries relevant here, the most general
metric in the Fefferman-Graham coordinates takes the form
\eq
ds^2=\f{1}{z^2} \left( -e^{a(z,\tau)} d\tau^2+e^{b(z,\tau)} \tau^2
dy^2 +e^{c(z,\tau)} dx_\perp^2 \right) +\f{dz^2}{z^2}
\eqx
with three coefficient functions $a(z,\tau)$, $b(z,\tau)$ and
$c(z,\tau)$. The energy density defining the physics is simply related to the
boundary asymptotics of the $a(z,\tau)$ coefficient:
\eq
\label{e.epsa}
\eps(\tau) = - \lim_{z \to 0} \f{a(z,\tau)}{z^4}
\eqx

Of course it is too difficult to explicitly perform this construction
for arbitrary functions $\eps(\tau)$. What one does in practice is to
perform an expansion of $\eps(\tau)$ for (large) proper-times $\tau$
and determine one by one the subsequent terms in the expansion. In the
following section we will review the results obtained so far in
carrying out this program \cite{JP1,SJSIN,RJ}.

\section{Review of the dual geometry up to $\OO{\tau^{-\f{4}{3}}}$}

In \cite{JP1} it was shown that in order to study large proper-time
limit of the metric one is led to introduce a scaling variable
\eq
v=\f{z}{\tau^{\f{1}{3}}}
\eqx
and take the limit $\tau \to \infty$ with $v$ fixed. In order to study
subleading terms in the metric we have to perform an expansion around
this limt.

Let us now expand the metric coefficient functions $a(z,\tau)$,
$b(z,\tau)$ and $c(z,\tau)$ in a series of the following form:
\eq
\label{e.aexp}
%a(z,\tau)=\sum_{i=0} a_i(v) \f{1}{\tau^{\f{2i}{3}}}
a(z,\tau)=a_0(v)+a_1(v) \f{1}{\tau^{\f{2}{3}}} +a_2(v)
\f{1}{\tau^{\f{4}{3}}} +a_3(v) \f{1}{\tau^2} +\ldots
\eqx
and similar expressions for the other coefficients. The motivation for
choosing such specific powers of $\tau$ is twofold. Firstly, on the
gauge theory side once we have viscous hydrodynamics with $\eta
\propto T^3$, the asymptotic expansion of the energy density
$\eps(\tau)$ exactly corresponds to the above decomposition of the
metric coefficients\footnote{Recall that $\eps(\tau)$ is extracted
  from minus the coefficient of $z^4$ in $a(z,\tau)$.} (\ref{e.aexp}).  
Secondly, one can show directly from the gravity side without assuming
anything on viscous dynamics in gauge theory, that a correction at
order $1/\tau^{\f{2}{3}}$ has to occur. Namely suppose that we start
from the leading order solution with a {\em generic} first correction
\eq
a(z,\tau)=a(v)+a_r(v) \f{1}{\tau^r}+\ldots
\eqx
Then we find that the square of the Riemann tensor is nonsingular at
the leading and $1/\tau^r$ orders but will {\em always} have a singularity at
order $1/\tau^{\f{4}{3}}$. There will also be a singularity at order
$1/\tau^{2r}$. The only possibility of obtaining a nonsingular
geometry is that $2r=\f{4}{3}$ and that these two singularities cancel,
which is indeed what occurs. This fixes the power of the first
subleading correction to be $1/\tau^{\f{2}{3}}$ which is exactly what
is expected for the form of corrections due to shear viscosity with
the coefficient behaving like $\eta \propto T^3$.
 
The procedure is now to insert the expansion (\ref{e.aexp}) into Einstein's
equations and solve them order by order. At each order a new
integration constant (free parameter) will occur and we will determine
it by requiring that a similar expansion of the square of the Riemann
tensor will be nonsingular i.e.
\eq
\rsq \equiv R_{\mu\nu\al\bt}R^{\mu\nu\al\bt} =   R_0(v)+R_1(v)
\f{1}{\tau^{\f{2}{3}}} +R_2(v) \f{1}{\tau^{\f{4}{3}}} +R_3(v)
\f{1}{\tau^2} +\ldots 
\eqx
with all $R_i(v)$ being nonsingular.

In order to present the solutions it turns out to be convenient to define
\eq
\btl_i(v) \equiv b_i(v)+2c_i(v)
\eqx

The leading order solution found in \cite{JP1} is
\eqn
a_0(v) &=& \log\f{(1-v^4/3)^2}{1+v^4/3} \nonumber \\
\btl_0(v) &=& \log(1+v^4/3)^3 \nonumber \\
c_0(v) &=& \log(1+v^4/3)
\eqnx 
and the resulting $\rsq$ coefficient is nonsingular:
\eq
R_0=\f{8(5v^{16}+60 v^{12}+1566 v^8+540v^4+405)}{(3+v^4)^4}
\eqx
From this expression, using the similarity with the static
black hole metric we may read off the temperature from the position of
the horizon to obtain at this order:
\eq
\label{e.T}
T(\tau)=\f{\sqrt{2}}{3^{\f{1}{4}} \pi \tau^{\f{1}{3}}}
\eqx
We will use this expression later in the paper.
The first subleading correction was found in \cite{SJSIN} and reads
in our conventions
\eqn
a_1(v) &=& 2\eta_0 \f{(9+v^4)v^4}{9-v^8} \nonumber\\
\btl_1(v) &=& -6 \eta_0 \f{v^4}{3+v^4} \nonumber \\
c_1(v) &=& -2\eta_0 \f{v^4}{3+v^4} -\eta_0 \log \f{3-v^4}{3+v^4}
\eqnx
where $\eta_0$ is an undetermined integration constant (which has the
physical interpretation as the coefficient of shear
viscosity). However it turns out that $\eta_0$ is undetermined at this
order from nonsingularity of $\rsq$ since
\eq
\label{e.rone}
R_1=\f{41472 (v^4-3)v^8}{(3+v^4)^5} \cdot \eta_0
\eqx
and is nonsingular for {\em any} value of $\eta_0$.
Therefore in order to fix $\eta_0$ one has to go one order
higher. This was done in \cite{RJ} with the result
\eqn
a_2(v) &=& \f{(9+5v^4)v^2}{6(9-v^8)} -C \f{(9+v^4)v^4}{36(9-v^8)} +
\eta_0^2 \f{(-1053-171v^4+9v^8+7v^{12})v^4}{3(9-v^8)^2}+
\f{1}{4\sqrt{3}} \log \f{\sqrt{3}-v^2}{\sqrt{3}+v^2}-
\f{3}{2} \eta_0^2 \log \f{3-v^4}{3+v^4} \nonumber \\
\btl_2(v) &=& \f{v^2}{2(3+v^4)} +C\f{v^4}{12(3+v^4)} +\eta_0^2
\f{(39+7v^4)v^4}{(3+v^4)^2} +
\f{1}{4\sqrt{3}} \log \f{\sqrt{3}-v^2}{\sqrt{3}+v^2}+
\f{3}{2} \eta_0^2 \log \f{3-v^4}{3+v^4} \nonumber \\
c_2(v) &=& -\f{\pi^2}{144\sqrt{3}} +\f{v^2(9+v^4)}{6(9-v^8)} +C
\f{v^4}{36(3+v^4)} -\eta_0^2 \f{(-9+54v^4+7v^8)v^4}{3(3+v^4)(9-v^8)} +
\f{1}{4\sqrt{3}} \log \f{\sqrt{3}-v^2}{\sqrt{3}+v^2}+ \nonumber \\
&&
+\f{1}{36} (C+66\eta_0^2) \log \f{3-v^4}{3+v^4}+
 \f{1}{12\sqrt{3}} \left( \log \f{\sqrt{3}-v^2}{\sqrt{3}+v^2}
\log\f{(\sqrt{3}-v^2)(\sqrt{3}+v^2)^3}{4(3+v^4)^2} -\li \left(-
\f{(\sqrt{3}-v^2)^2}{(\sqrt{3}+v^2)^2} \right)\right)
\eqnx
Here $C$ is a {\em new} free integration constant appearing at this
order and $\li$ is the dilogarithm function. A calculation of the
$\rsq$ coefficient at this order gives 
\eqn
R_2 &=& -\f{576 (v^4-3)v^8}{(3+v^4)^5} C+\f{6912(5v^{24}-60v^{20} +2313
v^{16}  -6912 v^{12}+26487 v^8-18468 v^4+13851)v^8 \eta_0^2}{(3-v^4)^4
(3+v^4)^6} - \nonumber\\
&& -\f{4608(5 v^{16}+6 v^{12}+162 v^8 +54 v^4+405) v^{10}}{(3-v^4)^4
  (3+v^4)^5} 
\eqnx
We see that there is a fourth order pole singularity. It may be
cancelled when the viscosity coefficient $\eta_0$ takes the value
\cite{RJ}
\eq
\label{e.eta0}
\eta_0=\f{1}{2^{\f{1}{2}} 3^{\f{3}{4}}}
\eqx
which reproduces, in the boost-invariant expanding setup, the exact
viscosity coefficient of $\nn=4$ SYM calculated in the static case in
\cite{son}. 

We see however, that at this order the new integration constant $C$ is
still undetermined. By analogy with the case of $\eta_0$ we may expect
that it will be fixed at the next order of the series
expansion. Before we proceed to do this let us first discuss the physical
interpretation of this coefficient.

\subsection*{The physical interpretation of $C$}

The energy density extracted from the metric expanded in the scaling
limit up to $\OO{\tau^{-\f{4}{3}}}$ through (\ref{e.epsa}) is
given by
\eq
\label{e.epstau}
\eps(\tau)=\left(\f{N_c^2}{2\pi^2} \right) \cdot
\f{1}{\tau^{\f{4}{3}}} \left\{1 -\f{2\eta_0}{\tau^{\f{2}{3}}}+ \left(
\f{10}{3} \eta_0^2+\f{C}{36} \right) \f{1}{\tau^{\f{4}{3}}} + \ldots
\right\}
\eqx
Let us now suppose that this behaviour can be described by the
equations of second order viscous hydrodynamics
(\ref{e.i})-(\ref{e.ii}) with some parameters $\eta$ and
$\tau_\Pi$. We will use this hypothesis to extract these
parameters\footnote{Since viscosity is known, the new information is
  the relaxation time $\tau_\Pi$.} which may then be used in
hydrodynamic simulations for generic 3+1 evolving plasma systems. 

For ease of computation we note that equations
(\ref{e.i})-(\ref{e.ii}) can be divided out by the factor
$N_c^2/(2\pi^2)$, thus we can drop this factor from $\eps$, $\eta$ and
$\Phi$. Let us now insert $\Phi$ from (\ref{e.i}) into (\ref{e.ii})
and obtain a differential equation expressed sorely in terms of
$\eps(\tau)$ and the {\em numerical} coefficients $A$ and $r$. 

We may now extract these numerical coefficients
by plugging in the energy density from the AdS/CFT calculation
(\ref{e.epstau}) into the resulting equation.

The result is
\eqn
A &=& \eta_0 \\
r &=& \f{-C-66\eta_0^2}{324 \eta_0^2}
\eqnx
$A$ is precisely equal to $\eta_0$ which is not unexpected since in
the leading order $\eps(\tau)=1/\tau^{\f{4}{3}}$ and hence
$\eta=\eta_0/\tau$ may be written as $\eta=\eta_0 \eps^{\f{3}{4}}$.
The value of $\eta_0$ is determined from nonsingularity to this order
\cite{RJ} to be given by (\ref{e.eta0}). The new information is $r$
defining the relaxation time $\tau_\Pi$. Plugging in the value
(\ref{e.eta0}) for $\eta_0$ we may express $r$ in terms of $C$:
\eq
\label{e.r}
r=-\f{11}{54}-\f{C}{18 \sqrt{3}}
\eqx  
Hence the relaxation time $\tau_\Pi$ will be found once we know
$C$ (we note that thus we must have $C<0$). Again, as was the case
with $\eta_0$ one has to go one order higher in order to determine the
value of the coefficient. We will proceed to do it in the following section. 

\section{Determination of $C$ and the relaxation time $\tau_\Pi$}

It is straightforward to obtain equations for the third order metric
coefficients $a_3(v)$, $b_3(v)$ and $c_3(v)$ using a computer algebra
system, albeit these equations appear to be prohibitively
complex at first sight.

However one can notice that the ${}_{zz}$ component of the
Einstein equations $E_{zz}=0$ (\ref{e.einst}) gives an equation just
for the combination
\eq
d'_3(z) \equiv a'_3(v)+\btl'_3(v) \equiv a'_3(v)+b'_3(v)+2c'_3(v)
\eqx 
where the prime stands for the ordinary derivative.
Then expressing $b'_3(v)$ in terms of the
other functions and plugging the result into the ${}_{\tau\tau}$
component $E_{\tau\tau}=0$ of Einstein equation yields an equation
purely for the derivative $a'_3(v)$. So at this stage we have analytic
expressions \cite{sourcefile} for $a'_3(v)$ and (say) $\btl'_3(v)$.

Fortunately, the expression for the $\OO{\tau^{-2}}$ coefficient $R_3$
of $\rsq$ can be expressed sorely in terms of $a'_3(v)$, $\btl'_3(v)$
and their derivatives. The expression is quite lengthy so we do not
give it here \cite{sourcefile}. Substituting $a_3'(v)$ and
$\btl'_3(v)$ into $R_3$, we find again that there is a fourth order
pole in $R_3$ at $v=3^{\f{1}{4}}$ which is canceled exactly when 
\eq
C=\f{6\log 2 -17}{\sqrt{3}}
\eqx
For this value the poles of lower orders are also canceled. However a
new feature arises here, namely there is a leftover {\em logarithmic}
singularity:
\eq
\label{e.riiilog}
{R_3}_{\| C=\f{6\log 2-17}{\sqrt{3}}} =finite +8 \cdot 2^{\f{1}{2}}
3^{\f{3}{4}}\log(3-v^4) 
\eqx
We will show in the next section that if we turn on the dilaton this
singularity may be canceled without modifying the value of $C$
determined above. Let us now determine the coefficient of the
relaxation time $r$ from (\ref{e.r}). We find that
\eq
\label{e.rfin}
r= \f{1}{9}(1-\log 2) \sim 0.034 \sim \f{1}{30}
\eqx
which shows that the relaxation time of $\nn=4$ SYM at strong coupling
is much smaller than one would expect. In particular the second order
viscous hydrodynamics seem to be much closer to the first order
behaviour. If we express the relaxation time in terms of the proper
time and then relate it to the temperature through the leading order
expression (\ref{e.T}), we obtain
\eq
\label{e.taupifin}
\tau_\Pi=\f{1-\log 2}{6\pi T}
\eqx

\section{The dilaton and $\tr F^2$}

In this section we will show how to cancel the remaining logarithmic
divergence in $R_3$ by turning on the dilaton field. On the gauge
theory side, this means that we generate a nonzero expectation value
for $\tr F^2$ i.e. electric and magnetic modes are no longer
equilibrated. Physically this may well happen since we are considering
a regime where dissipative effects are important.

If we include the dilaton $\phi$, the equation of motion become those
of a coupled Einstein-dilaton system which read (in the Einstein
frame):
\eqn
\label{e.einstdil}
R_{\mu\nu}+4g_{\mu\nu} &=& \f{1}{2} \partial_\mu \phi \partial_\nu
\phi \\
\partial_\mu \left(\sqrt{g} g^{\mu\nu} \partial_\nu \phi \right) &=& 0
\eqnx
In analogy to the metric coefficients (\ref{e.aexp}) we would like to
make a large $\tau$ expansion of the dilaton. This requires a
judicious choice of the powers of $\tau$ appearing in that
expansion. We fix those powers by the following argument.

Suppose that $\phi(z,\tau) \sim \tilde{\phi}(v)
\tau^{-r}$. Then one can check using (\ref{e.einstdil}) that the
dilaton source term will contribute to the metric coefficients at
order $\tau^{-2r}$. The above considerations suggest the following
expansion of the dilaton field:
\eq
\label{e.dilexp}
\phi(z,\tau)=\sum_{i=1} \phi_i(v) \f{1}{\tau^{\f{i}{3}}}
\eqx
The $i=0$ component is absent as has already been checked in
\cite{BAK}. 

If a spacetime has a nontrivial dilaton profile, in string theory one
has two distinguished metrics -- the Einstein frame metric considered
above, and the string frame metric defined as
\eq
g_{\mu\nu}^{string} =e^{\f{1}{2} \phi} g_{\mu\nu}
\eqx 
which is the natural metric from the point of view of {\em strings}
propagating in the spacetime. Therefore a natural question
arises to which of the two metrics we should apply the nonsingularity
criterion. A natural guess would be that $\rsq(g_{\mu\nu}^{string})$
should be nonsingular, we will however for the moment keep both
possibilities open.

Let us now go back to the expansion (\ref{e.dilexp}). The only terms
that would contribute to the metric coefficients\footnote{And
  similarly for $b_i(v)$ and $c_i(v)$.}  $a_1(v)$, $a_2(v)$
and $a_3(v)$ are $\phi_1(v)$, $\phi_2(v)$ and $\phi_3(v)$. So only
these terms would influence $\rsq$ in the Einstein frame to the order
considered in the previous sections. If on the other hand we would
require the nonsingularity of $\rsq$ in the {\em string} frame, then
we must consider in addition $\phi_4(v)$, $\phi_5(v)$ and $\phi_6(v)$.

Finally let us give one gauge theoretical interpretation of a
nontrivial dilaton profile. It might give rise to an expectation value
for $\tr F^2$ through \cite{vijay,wittenkleb}
\eq
\label{e.trfsqform}
\f{1}{4g^2_{YM}}\cor{\tr F^2}=\f{N^2}{2\pi^2} \cdot \lim_{z \to 0}
\f{\phi(z,\tau)}{z^4} 
\eqx

In the following we will consider various possible leading behaviours of the
dilaton according to the expansion (\ref{e.dilexp}) and consider
$\rsq$ in string frame (and in Einstein frame where relevant)

\bigskip
 
\noindent{\bf $\phi$ at $\OO{\tau^{-1/3}}$.} At this order the leading
behaviour of $\phi$ can be easily obtained as
\eq
\label{e.leading}
\phi_1(v)=k \log \f{3-v^4}{3+v^4}
\eqx
In the string frame $\rsq_{string}$ will have a new contribution at
order $\tau^{-1/3}$. This contibution has a piece which is
proportional to $k\log(3-v^4)$. Therefore we have to set $k=0$ and
hence $\phi_1(v)=0$. Let us note that the
natural appearance of a logarithmic divergence gives a hope that
indeed the dilaton might cancel the leftover divergence in
(\ref{e.riiilog}). We will see below that this indeed happens. 

In the Einstein frame the calculations are more complicated. $\phi_1$ will
modify the metric coefficients $a_1(v)$, $b_1(v)$ and $c_1(v)$. Then
the Einstein frame $\rsq$ will also get modified and (\ref{e.rone})
will be replaced by
\eq
R_1=\f{41472 (v^4-3)v^8}{(3+v^4)^5} \cdot \eta_0 -\f{1152(13v^{12}-99
  v^8+27 v^4+27) v^8}{(3+v^4)^5 (3-v^4)^2} \cdot k^2
\eqx
Requiring nonsingularity again gives $k=0$ and hence $\phi_1(v)=0$.

\bigskip
 
\noindent{\bf $\phi$ at $\OO{\tau^{-2/3}}$.} The leading part
$\phi_2(v)$ has a functional form identical to
(\ref{e.leading}). Again the string frame $\rsq_{string}$ will have a
logarithmic singularity but now at order $\tau^{-2/3}$, hence
$\phi_2(v)$ has to vanish at this order if we assume string frame
nonsingularity.

In the Einstein frame, the coefficients $a_2(v)$, $b_2(v)$ and
$c_2(v)$ will get modified. $\rsq$ will still have a fourth order pole
at $\OO{\tau^{-4/3}}$ 
which will be canceled by taking (\ref{e.eta0}) but an additional
second order pole proportional to $k$ will persist. Hence also
requiring nonsingularity in the Einstein frame will lead to $\phi_2(v)=0$. 

\bigskip
 
\noindent{\bf $\phi$ at $\OO{\tau^{-1}}$.} Here again in the string
frame one has a logarithmic singularity hence $\phi_3(v)=0$. The
calculations in Einstein frame are very tedious but also rule out a
nonvanishing $\phi_3(v)$.

\bigskip
 
\noindent{\bf $\phi$ at $\OO{\tau^{-4/3}}$.} At this order, the metric
coefficients are not modified, hence it makes sense to consider only
the string frame $\rsq_{string}$. A logarithmic singularity which
appears at order $\tau^{-4/3}$ requires $\phi_4(v)=0$.

\bigskip
 
\noindent{\bf $\phi$ at $\OO{\tau^{-5/3}}$.} Again a logarithmic
singularity requires $\phi_5(v)=0$.

\bigskip
 
\noindent{\bf $\phi$ at $\OO{\tau^{-2}}$.} This is the relevant order
at which a cancelation of the logarithmic singularity may occur.
At this order $\phi_6$ has again the form
\eq
\label{e.dilaton}
\phi_6(v)=k \log \f{3-v^4}{3+v^4}
\eqx
The string frame $\rsq_{string}$ at $\OO{\tau^{-2}}$ will get modified
by
\eq
R_3^{string}=R_3-k\cdot \f{8(5v^{16}+60v^{12} +1566 v^8+ 540
  v^4 +405)}{(3+v^4)^4} \log \f{3-v^4}{3+v^4}
\eqx
where $R_3$ is the Einstein frame coefficient calculated earlier.
Therefore we see that by a suitable choice of $k$ we may exactly cancel
the leftover logarithmic singularity without changing the value of $C$
obtained above which came from canceling a fourth order pole. Indeed
performing the expansion of $R_3^{string}$ at $v=3^{1/4}$ we get
\eq
R_3^{string}=finite+\left(8 \cdot 2^{\f{1}{2}} 3^{\f{3}{4}}  -112 k
\right)\log(3-v^4) 
\eqx
which gives
\eq
\phi_6(v)=\f{3^{\f{3}{4}}}{7\sqrt{2}} \log \f{3-v^4}{3+v^4}
\eqx
Now we may evaluate the resulting expectation value of $\cor{\tr F^2}$
from (\ref{e.trfsqform}):
\eq
\label{e.trfsqfin}
\f{1}{4g^2_{YM}}\cor{\tr F^2}= \f{N^2}{2\pi^2} \cdot
  -\f{\sqrt{2}}{7\cdot 3^{\f{1}{4}}} \cdot \f{1}{\tau^\f{10}{3}} 
\eqx
We see that this expectation value is {\em negative} which signifies
that electric modes become dominant. 

\section{Discussion}

In this paper we have considered the subasymptotic proper-time
evolution of an infinite expanding boost-invariant plasma in $\nn=4$
SYM at strong coupling. We have used the AdS/CFT correspondence to
construct dual geometries and used the criterion that the unique
nonsingular geometry corresponds to the physical evolution of the
plasma. Then the proper-time dependence of the energy density
$\eps(\tau)$ can be
read off from the form of the 5-dimensional metric.

We have found subleading terms of $\eps(\tau)$ which are sensitive to
the relaxation time $\tau_\Pi$ appearing, in addition to the shear
viscosity $\eta$, as a new element in second order viscous
hydrodynamics. This served to determine the relaxation time
(\ref{e.rfin})-(\ref{e.taupifin}) which
turned out to be about thirty times shorter than the one expected from
Boltzmann kinetic theory estimates (\ref{e.taubol}). Thus second
order hydrodynamics appears to be much closer to the ordinary first
order formalism.

In addition we have found that canceling a remaining logarithmic
divergence requires turning on the dilaton field\footnote{At least
  turning on the dilaton cancels the singularity in string frame.}. A
nonvanishing dilaton field in turn implies a nonzero expectation value
(\ref{e.trfsqfin}) for $\tr F^2$ with the proper-time scaling
$\tau^{-10/3}$. This means that electric and magnetic modes are not
exactly equilibrated. 
%A similar difference of electric and magnetic
%modes appears to be generic for plasma instabilities (albeit at weak
%coupling) \cite{plasma}.

It would be very interesting to understand from a microscopic
perspective the very short relaxation time. In particular it would be
interesting to understand the weak coupling corrections to the
Boltzmann value $\taubol$. Another intriguing question would be to
estimate by other means the $\tau$ dependence of $\cor{\tr F^2}$.
On the strong coupling side it would be interesting to understand the
dual metric from the point of view of dynamical horizons in general
relativity \cite{Ashtekar} and analyze the relevant thermodynamics.      

\bigskip

\noindent{}{\bf Acknowledgments. } RJ would like to thank Dongsu Bak,
 Robi Peschanski and participants of the GGI workshop `High density
 QCD' for discussions and comments. RJ would like to thank the Galileo
 Galilei Institute for Theoretical Physics for the hospitality and the
 INFN for partial support during the completion of this work. We thank
 Yuri Kovchegov for pointing out an error in the previous version.
 This work has been supported in part by Polish
Ministry of Science and Information Society Technologies grants
1P03B02427 (2004-2007), 1P03B04029 (2005-2008), RTN network ENRAGE
MRTN-CT-2004-005616, and the Marie Curie Actions Transfer of
Knowledge project COCOS (contract MTKD-CT-2004-517186).

\end{document}